\documentclass[11pt,twoside]{article}

\usepackage{asp2004}
\usepackage{epsf}
\usepackage{psfig}
\usepackage{lscape}

\begin{document}
\title{Spectropolarimetry of cool stars} 
\author{P. Petit}  
\affil{Observatoire Midi-Pyr\'en\'ees, 14 avenue Edouard Belin, 31400 Toulouse, France}   

\begin{abstract} 
In recent years, the development of spectropolarimetric techniques deeply modified our knowledge of stellar magnetism. In the case of solar-type stars, the challenge is to measure a geometrically complex field and determine its evolution over very different time frames. In this article, I summarize some important observational results obtained in this field over the last two decades and detail what they tell us about the dynamo processes that orchestrate the activity of cool stars. I also discuss what we learn from such observations about the ability of magnetic fields to affect the formation and evolution of Sun-like stars. Finally, I evoke promising directions to be explored in the coming years, thanks to the advent of a new generation of instruments specifically designed to progress in this domain.
\end{abstract}

\section{Introduction}  

Studying the magnetism of Sun-like stars is both an exciting and challenging task. The observation of stars resembling the Sun by their extended convective envelope (i.e. with a stellar mass lower than about $1.5 M_\odot$) brings the opportunity to investigate how stellar magnetic activity phenomena vary over an array of stellar parameters (age, mass, rotation rate, binarity...). The interpretation of observations requires the transposition of solar dynamo models for various internal structures. This transposition is an opportunity to study how some activity phenomena observed in the Sun can be excited or inhibited in other situations, a promising option to progress on open questions remaining in our understanding of the solar dynamo. 

This scientific interest is however coming along with challenging technical difficulties, the most obvious being the lack of direct spatial resolution on stellar surfaces. This situation imposed the development of a completely new set of dedicated tools, from spectropolarimetric instrumentation to tomographic imaging techniques.  

\section{Multi-line extraction of Zeeman signatures}

\begin{figure}[t]
   \centering
   \includegraphics{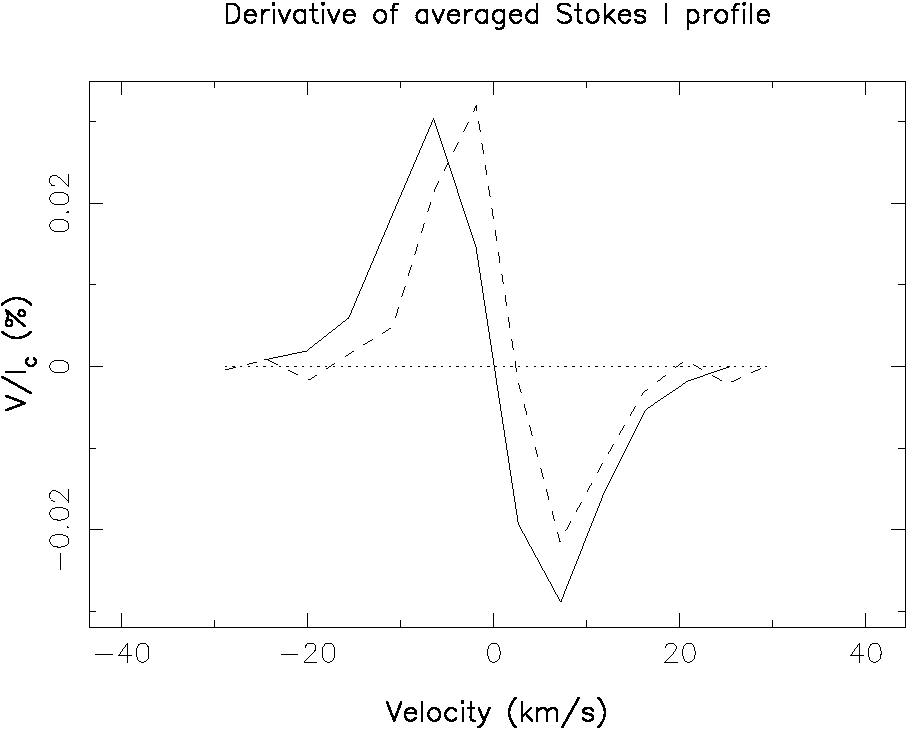}
   \caption{Derivative of the Stokes I LSD profile of the G8 dwarf $\xi$ Boo (full line), plotted together with the corresponding Stokes V profile. The asymmetry of the Stokes V profile appears as a larger area and amplitude of its blue lobe. An apparent velocity shift between both curves is also observed, probably linked to a large-scale toroidal component of the surface magnetic field (Petit et al. 2005).}
   \label{fig:xiboo}
\end{figure}

The easiest option to measure a magnetic field in a stellar photosphere is to analyze the ability of the field to affect spectral lines through the Zeeman effect, i.e. either measure the broadening or the polarization of magnetic lines. The first effect is useful for targets for which the line-broadening is not dominated by non-magnetic effects (in particular rotational broadening) and was investigated on a number of cool stars (e.g. Johns-Krull \& Valenti 1996, Ruedi et al. 1997).  In many cases, the line polarization provides a less ambiguous measurement technique (reliable even for rapidly rotating stars) and has the advantage to give some additional information about the field orientation. The intricate magnetic topology of solar-type stars (featuring at any time a pattern of mixed polarities on the visible stellar hemisphere) is however responsible for an important decrease of this signal through mutual cancellation of Zeeman signatures formed in neighboring regions. The consequence is that, in the vast majority of cases, only circular polarization (Stokes V) is detectable. The most active Sun-like stars produce a Stokes V signal amplitude lower than about $10^{-2} I_c$ (where $I_c$ is the continuum level), even for the most adequate spectral lines available in the optical domain (Donati et al. 1990). The signal amplitude decreases by approximately a factor of 100 for stars of intermediate activity level (Petit et al. 2005). The Sun itself, if it was observed as a star, would produce circular polarization at a level about 1000 times lower than the most active stars.

In this context, the noise level is often the main limitation to magnetic field measurements. This situation can be greatly improved by extracting simultaneously the polarized signal contained in a large number of magnetic lines. The multi-line code LSD (Least-Square Deconvolution) was developed with this aim by Donati et al. (1997). For a star of spectral type K1, and assuming that the spectrograph covers a spectral domain ranging from 370~nm to 1,000~nm (corresponding to the spectral window of ESPaDOnS, see Donati et al. in these proceedings), about 5,000 spectral lines are simultaneously recorded, yielding to the extraction of an average line profile with a noise level reduced by a factor of about 50 with respect to the raw spectrum. This technique is successfully used over a growing sample of stars, enabling field measurements at various locations in the HR diagram, over a range of evolutionary stages covering most of the evolution of Sun-like stars, from the pre-main sequence (e.g. Donati \& Collier Cameron 1997) to the sub-giant branch (e.g. Petit et al. 2004b). 

A growing range of surface temperature is also explored, or quite equivalently a growing range of depths of the convective zone. Available observations start with late F dwarfs (Marsden et al., these proceedings) possessing shallow envelopes. Recent studies by Donati et al. (2006) or Berdyugina et al. (these proceedings) now concentrate on the opposite side of the temperature scale, with field detections on fully convective M dwarfs, providing a first insight in a regime where stars do not possess a tachocline and where the dynamo processes at the origin of their field must be very different from the solar dynamo.    

Spectropolarimetric observations were concentrating at first on stars rotating 10 to 50 times faster than the Sun (Donati et al. 2003a), because their fast rotation ensures the onset of a very efficient dynamo, generating strong surface fields. Thanks to the improved accuracy of new instruments, possible targets now include much less active objects (Petit et al. 2005).  New spectropolarimeters like ESPaDOnS allow longitudinal field measurements with an accuracy better than about 1 Gauss for the brightest stars, which will allow one to measure background fields on stars no more active than our Sun. This progress in accuracy will result in a significant enlargement of the sample of observed stars in the coming years. 

Recent observations of the G8 dwarf $\xi$~Bootis~A (Petit et al. 2005 and Fig~\ref{fig:xiboo}) revealed a systematic asymmety of Stokes V profiles, with a larger amplitude and area of the blue lobe of the profiles. The level of asymmetry is similar to what is observed on the Sun (e.g. Pantellini, Solanki \& Stenflo 1988). The asymmetry of solar profiles is usually modeled by the combined use of velocity and magnetic gradients in magnetic elements (see Solanki 1993 for a review). The sign and level of asymmetry depend on the nature of magnetic regions in which the Zeeman signature is formed (faculae and network field are considered by Solanki 1989, sunspot penumbra by Solanki \& Montavon 1993). A dependence with the limb angle is also to be taken into account (Stenflo, Solanki \& Harvey 1987). In the future, a rigorous modeling of asymmetric profiles may help to constrain thermal and dynamical properties of magnetic elements producing the measured polarization.

\section{Magnetic mapping of stellar photospheres}

\begin{figure}[t]
   \centering
   \includegraphics{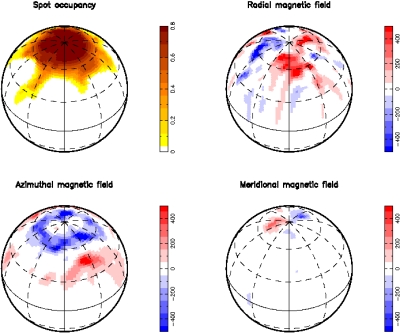}
   \caption{Photospheric magnetogram of the G5 sub-giant HD~199178, reconstructed by means of Zeeman-Doppler Imaging. The top-left panel represents the distribution of brightness inhomogeneities, while the other charts show all three components of the magnetic field, in spherical coordinates. The scale is given in Gauss (Petit et al. 2004b).}
   \label{fig:hd199178}
\end{figure}

The spatial distribution of photospheric magnetic fields can be reconstructed for fast rotating stars (with a significant rotational broadening of spectral lines) by means of tomographic inversion techniques very similar to Doppler imaging (Vogt \& Penrod 1983), but based on the inversion of polarized light and therefore called Zeeman-Doppler Imaging (Semel 1989, Donati \& Brown 1997). For cool stars, circularly polarized signal is generally used alone. The location of magnetic regions is obtained very similarly to that of star-spots in classical Doppler mapping. Some information about the orientation of field lines can also be determined by following the distortion of Zeeman signatures during the transit of magnetic regions over the stellar disc. If circular polarization is used alone, the inversion problem is ill-posed and the reconstructed orientation of the field is suffering from a limited crosstalk between different field components (in particular between the radial and meridional field). This effect is however mainly limited to stars with low inclination angle. In principle, this problem can be reduced by using simultaneously circularly and linearly polarized light for image reconstruction (Piskunov 2006). The detection of Zeeman signatures in linear polarization, out of reach of previous spectropolarimeters, is now possible with ESPaDOnS for a few active stars. 

In its first version, the ZDI algorithm (decomposing the stellar surface into a grid of independent pixels) was not able to reconstruct large-scale fields, like global dipoles. A new version of the code, assuming a spherical harmonic decomposition of the reconstructed field, in now solving this problem (Donati 2001). Several ZDI maps of cool spots and surface magnetic field have been reconstructed for a sample of active fast rotators of spectral types G and K. Some key results of this long-term study (some stars have been monitored over a timescale of one decade) are summarized hereafter.

While cool spots only cover about 10$^{-4}$ of the surface of the Sun at solar maximum, many observations of fast rotators demonstrate that spot coverage usually reaches several percents of the surface on these extremely active objects. The important spottedness is not only due to a larger number of spots, but also to the large typical size of ``giant'' individual structures (Fig. \ref{fig:hd199178}). The spatial distribution of cool spots also differs from the solar case, fast rotators generally hosting spots at all latitudes (and sometimes mostly in the polar region), while sun-spots are confined at low and intermediate latitudes. Taking into account the effect of rotation (which tends to deflect the emergence of flux tubes toward the pole in fast rotators, Schuessler et al. 1996), observations of spots over a large range of latitudes indicate that active structures may be formed in several layers (on in the totality) of the convective envelope (spots originating from the tachocline, as in the Sun, could not emerge at low latitude in rapidly-rotating stars), suggesting that the underlying dynamo may be active in the whole convective zone.

Magnetic regions observed on fast rotators are not always spatially associated with the distribution of cool spots. Large structures hosting an homogeneous field are observed outside cool regions, and could be associated to small brightness inhomogeneities unresolved by the imaging procedure. Also, if a magnetic field is usually detected in the largest star-spots (in particular in the prominent polar spots observed on several fast rotators), the measured field is generally weaker than expected if the giant spots were large analogues of solar spots. A first reason is that the relative darkness of cool spots results in a smaller amount of polarized signal and therefore in a more difficult detection of the field inside the spot. A second reason is that giant spots may in fact be constituted of several smaller spots of opposite polarities, resulting in a mutual cancellation of the polarized signal. 

The most intriguing magnetic structures observed in very active stars are large areas where the field is mostly azimuthal (the azimuthal component can even store most of the magnetic energy in some cases). The azimuthal field is sometimes organized in rings encircling the pole at various latitudes. The origin of such toroidal fields is still uncertain, but such features may be related to the large-scale toroidal component of a distributed dynamo. In fully convective stars, this component vanishes (Donati et al. 2006), suggesting that its presence is linked to the existence of a radiative core (disappearing below 0.3 M$_\odot$).

For slowly-rotating stars, the mutual cancellation of Zeeman signatures between neighboring magnetic regions is important, since all magnetic spots of the visible hemisphere possess similar line-of-sight velocities, so that the relative Doppler shifts are not sufficient to allow a partial disentangling of their spectral contributions. In such situation, individual active regions remain unresolved, but low-order components of the surface field (analogous to the {\it background} solar dipole showing up outside the belts of active regions) is still detectable. This component is investigated by Petit et al. (2005) for the G8 dwarf $\xi$~Bootis~A. On this moderately active star, the structure of the global field is not a simple dipole as in the Sun. The modeling proposed by Petit et al. suggests the presence of a significant toroidal component, with field strengths of 40 and 100 Gauss for the dipole and the azimuthal field, respectively.  

\section{Measurement of large-scale surface flows}

\begin{figure}[t]
   \centering
   \includegraphics{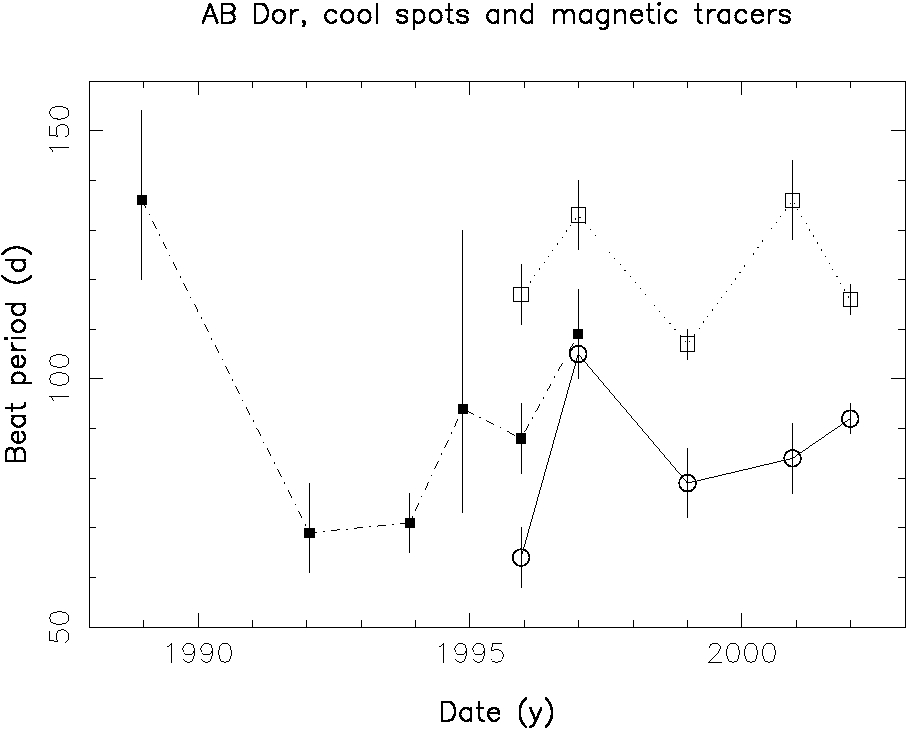}
   \caption{Differential rotation of AB~Dor as a function of time. The beat period is the time for the equator to lap the pole by one complete rotation cycle. Black squares represent measurements of Collier Cameron \& Donati (2002), using a direct spot tracking method. White symbols represent estimates performed by means of parametric imaging (Donati et al. 2003b). White squares correspond to cool spots, white circles to magnetic tracers.}
   \label{fig:hd199178}
\end{figure}

The short-term temporal evolution of surface structures can be analyzed with Zeeman-Doppler Imaging. Measurements of surface differential rotation can in particular be performed, using cool spots or magnetic regions as tracers of the large-scale surface flows. A solar-like surface shear (the equator rotating faster than the pole) has been detected on several stars (see, e.g., Donati, Collier Cameron \& Petit 2003b, Petit et al. 2004a). An exciting result is also the recent detection, on the young dwarf AB~Dor, of secular fluctuations of differential rotation (Collier Cameron \& Donati 2002, Donati et al. 2003b), about 40 times stronger in amplitude than fluctuations reported for the Sun (Howe et al. 2000, Vorontsov et al. 2002). This observation may unveil, for the first time on a star other than the Sun, the feedback effect of magnetic fields on the dynamics of convective zones (through Lorentz forces) during stellar activity cycles (Applegate 1992).

\section{Exploring the vertical structure of stellar magnetic fields}

\begin{figure}[t]
   \centering
   \includegraphics{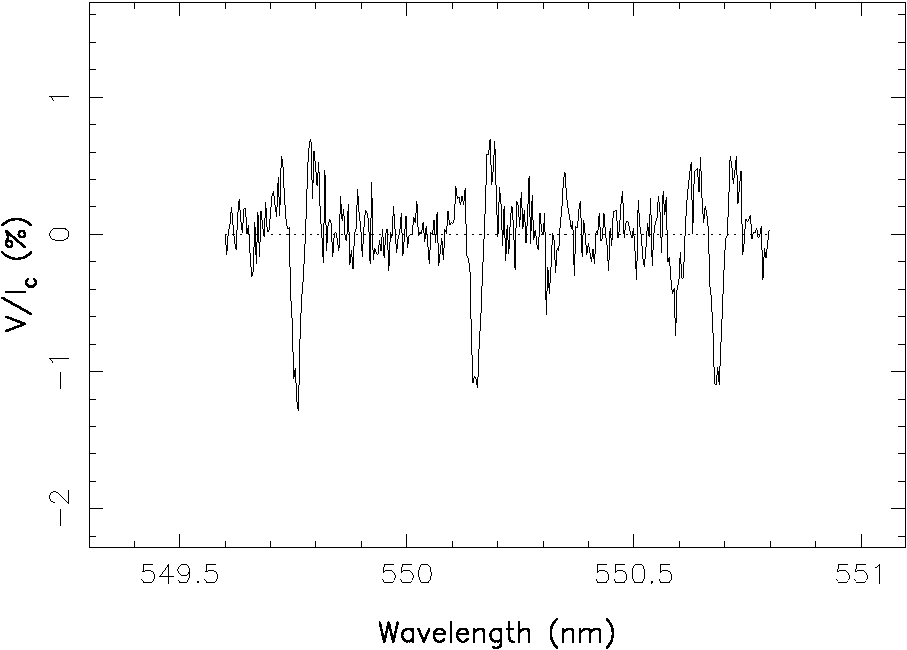}
   \includegraphics{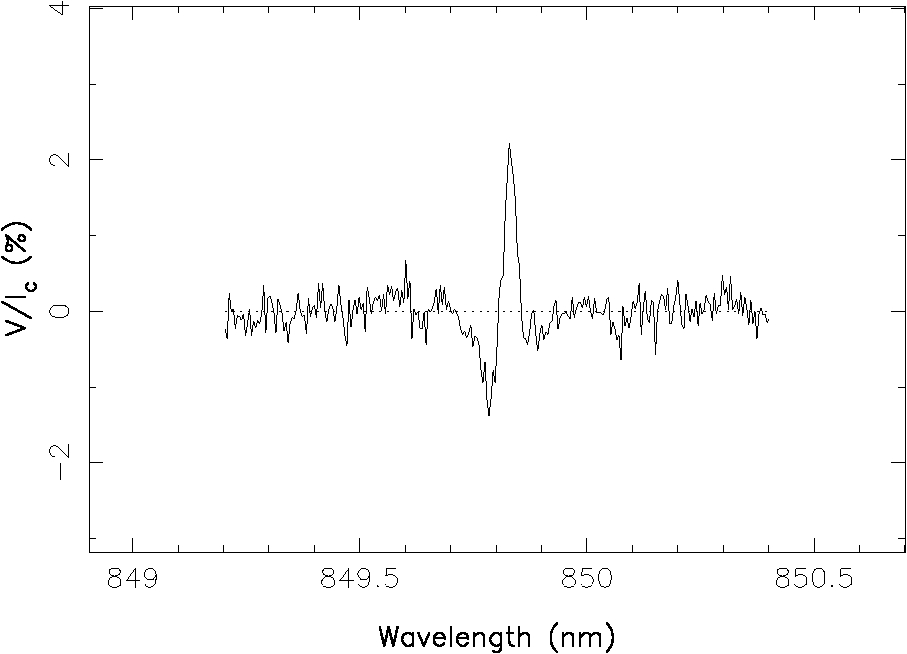}
   \caption{Examples of Stokes V line profiles of the K1 sub-giant II Peg, extracted from two spectral regions of a single ESPaDOnS spectrum. Upper frame~: three FeI lines around 550~nm. Lower frame~: the emission core of the CaII line at 849.8~nm.}
   \label{fig:lines}
\end{figure}

The collecting power of ESPaDOnS, combined with its excellent polarimetric accuracy, is sufficient to measure Zeeman signatures in individual lines of very active stars. The circular polarization signal is reaching a level of about 1\% for a few spectral lines with high Land\'e factors, for the K1 subgiant II~Peg (Fig. \ref{fig:lines}). The instrumental situation has greatly improved since the first successful attempt at detecting such signatures on a cool star (Donati et al. 1990). Today, the broad spectral domain of ESPaDOnS enables one to record a large number of magnetic lines in a single exposure, with tens of lines exhibiting a clear Stokes V signal. 

The available array of magnetic lines does not only include neutral metal lines. Several molecular bands are also recorded (in particular the magnetically sensitive TiO band around 705.5~nm), providing a probe of the coolest parts of the photospheres of G and K dwarfs, where they are mainly formed. Up to now, ZDI maps of such stars were reconstructed from atomic lines and suffered from a lack of sensitivity in the darkest active regions (because of the brightness contrast with the quiet photosphere). In particular, the molecular Zeeman effect is a promising tool to investigate the magnetic properties of the giant polar spots observed in many rapidly-rotating G and K stars (Berdyugina 2002). The first ESPaDOnS data set obtained with this aim suggests that the magnetic signal generated by such giant spots is weaker than expected if they were hosting an homogeneous field (Berdyugina et al. in these proceedings, Afram et al. in these proceedings). This observation gives therefore further support to the idea that their magnetic field may be an entangled pattern of mixed polarities. 

\begin{figure}[t]
  \centering
   \mbox{\includegraphics{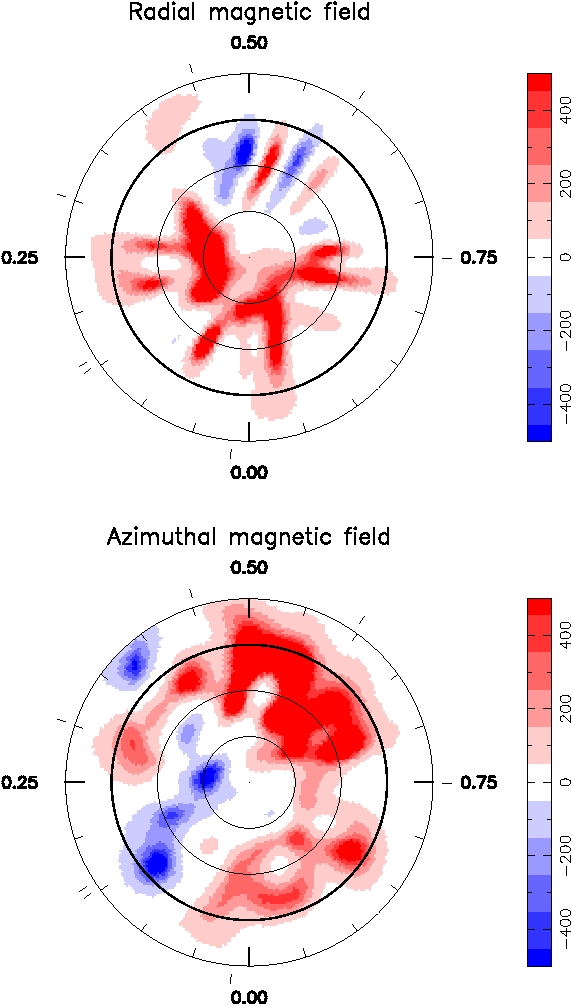} \hspace{20 mm} \includegraphics{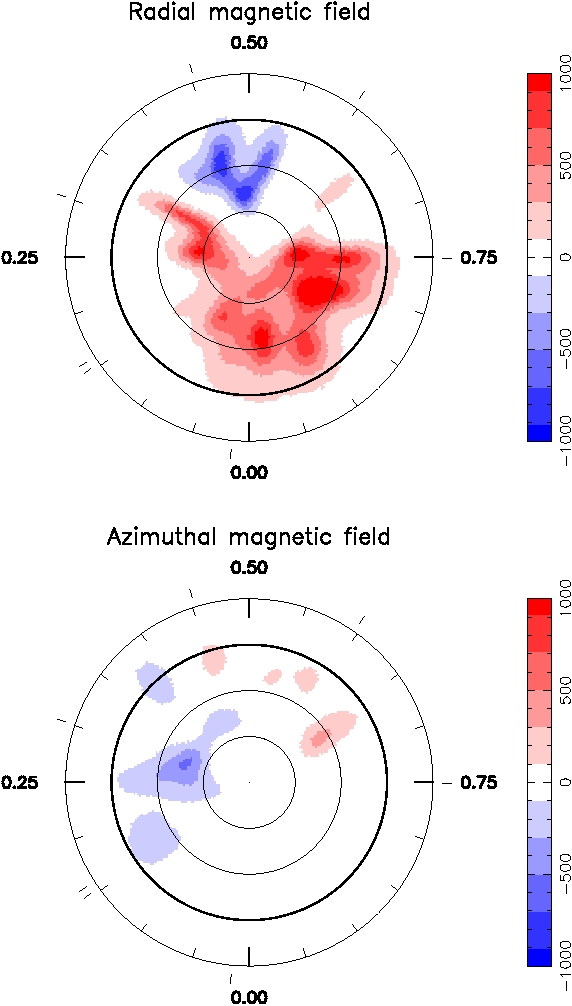}}
   \caption{Two magnetic maps of the K1 subgiant II Peg, in flattened polar view. The bold circle corresponds to the equator and the map extends down to latitude -30$^\circ$. Left panels (top and bottom)~: map reconstructed from photospheric lines. Right panels~: chromospheric map obtained from the three components of the CaII infrared triplet.}
   \label{fig:chromosphere}
\end{figure}

Zeeman signatures are also detected in the emission core of chromospheric lines for active K and M stars, in particular in all components of the CaII infrared triplet. It can be seen from Fig. \ref{fig:lines} that the sign of such polarized signal is opposite to that originating from neutral lines (as expected in the case of an emission component). However, the signatures  do not differ by their sign only, but also by their global shapes. This point is illustrated in Fig. \ref{fig:lines}, where the bluest (negative) lobe of the Zeeman signature recorded in the CaII line at 849.8~nm is deeper than the same lobe observed in each of the three FeI lines around 550~nm. The most likely interpretation is that variations in Zeeman signatures between different spectral lines reflect the different atmospheric layers in which the lines are formed, and therefore the different field geometries belonging to each layer. Using the technique of ZDI with individual lines (or groups of lines) brings the new opportunity to explore various layers of the stellar atmosphere, from the photosphere to the chromosphere. 

In Fig. \ref{fig:chromosphere}, we compare two magnetic maps of the K1 subgiant II~Peg. The first one is obtained from a time-series of LSD profiles computed from all 5,000 photospheric lines available in the spectra. The second map is reconstructed from another series of LSD profiles, but this time the mean profiles are extracted from the three components of the CaII infrared triplet (the profile of their emission cores, in which the Zeeman signatures are detected, is consistent with the rotational broadening observed in photospheric lines). The radial field component is quite well repeated in both maps, the main patches of radial field staying grossly at the same location, with a same polarity. The situation is very different for the azimuthal field component, which appears to be much weaker in the chromosphere than in the photosphere. From the magnetograms, we estimate that about 75\% of the magnetic energy is stored as horizontal magnetic field at photospheric level, while only 5\% of the total magnetic energy of the chromosphere is showing up in this component. We therefore conclude that azimuthal field regions are confined to the lowest layers of the atmosphere of II~Peg, the magnetic topology of the chromosphere being closer to a force-free field (similarly to the solar chromosphere, e.g. Metcalf et al. 1995). Magnetic chromospheric maps could therefore provide more adequate boundary conditions to coronal field extrapolations (Jardine, Collier Cameron \& Donati 2002) and help to obtain better magnetic models of stellar active coronae.

\section{Probing the magnetized environment of forming stars}

The formation of solar-type stars comes along with the temporary formation of an accretion disc (the T Tauri phase), surviving during a few million years, while the star is still in a purely convective state and contracting down its Hayashi track. The accretion of material from the disc onto the star is a crucial phenomenon that impacts the whole evolution of the star, with the determination of its future mass and angular momentum, as well as the potential formation of a planetary system. 

The accretion process is influenced by the magnetic field of T Tauri stars (their high activity level being witnessed through various indirect tracers like X rays emission, see e.g. Montmerle et al. 1993 for a review). The extended magnetosphere is expected to truncate the inner part of the disc and to channel accreted material onto the star along the magnetic lines, creating accretion columns and strong accretion shocks at the location of photospheric magnetic spots. The models of magnetospheric accretion are usually based on simple field geometries (in the absence of observational constraints), whereas recent simulations demonstrate that accretion processes can be deeply influenced by the actual magnetic geometry of the magnetosphere (Gregory et al. 2005).  Zeeman signatures have been detected on a limited sample of stars, through circular polarization of emission lines formed in accretion columns (in particular HeI at 587.6~nm). These observations reveal a typical field strength of several kG (Johns-Krull et al. 1999, Symington et al. 2005). More recently, a magnetic field of several hundred Gauss was detected in the protostellar accretion disc FU Ori (Donati et al. 2005). A modelling of the field structure indicates the presence of a significant azimuthal component, in agreement with models proposing a magneto-centrifugal launching of collimated jets.

\section{Future prospects}

The ESPaDOnS spectropolarimeter is now duplicated. NARVAL, the first clone, is expected to see its first light next summer at Observatoire du Pic du Midi. While ESPaDOnS is mostly used for short  programs, NARVAL will be partly dedicated to longer projects, involving dense rotational coverage necessary for tomographic imaging, or longer term monitoring for studies of stellar magnetic cycles. Another option would be the combined use of ESPaDOnS and NARVAL, ensuring almost continuous observation of stellar sources. This strategy will be useful for tomographic programs requiring a very dense rotational sampling of targets over several days, e.g. for analyses of global surface flows requiring extremely accurate surface reconstruction. 

On a longer term, a promising perspective is also the development of new instruments for spectropolarimetric studies in the near infrared domain (between 1 and 5 $\mu m$). Infrared observations would benefit from the easiest detection of the Zeeman effect at longer wavelengths. The study of ultra-cool dwarfs would also benefit from their higher flux in this spectral domain and from the number of molecular bands showing up in this wavelength range. Finally, the brightness contrast between spots and quiet photosphere of G/K dwarfs is lower at such wavelengths (with respect to the visible domain), and would allow precise studies of the magnetic properties of stellar spots.

\end{document}